\documentclass[10pt]{amsart}
\usepackage[dvips]{graphicx}

\usepackage{mathrsfs}
\usepackage{amssymb}
\usepackage{amsfonts}
\usepackage{latexsym}
\usepackage{amsthm}

\usepackage{graphicx}
\def\lb{\label}

\newcommand{\er}[1]{\textrm{(\ref{#1})}}

\begin{document}

\baselineskip14pt

\renewcommand{\theequation}{\arabic{section}.\arabic{equation}}
\theoremstyle{plain}
\newtheorem{theorem}{\bf Theorem}[section]
\newtheorem{lemma}[theorem]{\bf Lemma}
\newtheorem{corollary}[theorem]{\bf Corollary}
\newtheorem{proposition}[theorem]{\bf Proposition}
\newtheorem{definition}[theorem]{\bf Definition}
\newtheorem{remark}[theorem]{\it Remark}

\def\a{\alpha}  \def\cA{{\mathcal A}}     \def\bA{{\bf A}}  \def\mA{{\mathscr A}}
\def\b{\beta}   \def\cB{{\mathcal B}}     \def\bB{{\bf B}}  \def\mB{{\mathscr B}}
\def\g{\gamma}  \def\cC{{\mathcal C}}     \def\bC{{\bf C}}  \def\mC{{\mathscr C}}
\def\G{\Gamma}  \def\cD{{\mathcal D}}     \def\bD{{\bf D}}  \def\mD{{\mathscr D}}
\def\d{\delta}  \def\cE{{\mathcal E}}     \def\bE{{\bf E}}  \def\mE{{\mathscr E}}
\def\D{\Delta}  \def\cF{{\mathcal F}}     \def\bF{{\bf F}}  \def\mF{{\mathscr F}}
\def\c{\chi}    \def\cG{{\mathcal G}}     \def\bG{{\bf G}}  \def\mG{{\mathscr G}}
\def\z{\zeta}   \def\cH{{\mathcal H}}     \def\bH{{\bf H}}  \def\mH{{\mathscr H}}
\def\e{\eta}    \def\cI{{\mathcal I}}     \def\bI{{\bf I}}  \def\mI{{\mathscr I}}
\def\p{\psi}    \def\cJ{{\mathcal J}}     \def\bJ{{\bf J}}  \def\mJ{{\mathscr J}}
\def\vT{\Theta} \def\cK{{\mathcal K}}     \def\bK{{\bf K}}  \def\mK{{\mathscr K}}
\def\k{\kappa}  \def\cL{{\mathcal L}}     \def\bL{{\bf L}}  \def\mL{{\mathscr L}}
\def\l{\lambda} \def\cM{{\mathcal M}}     \def\bM{{\bf M}}  \def\mM{{\mathscr M}}
\def\L{\Lambda} \def\cN{{\mathcal N}}     \def\bN{{\bf N}}  \def\mN{{\mathscr N}}
\def\m{\mu}     \def\cO{{\mathcal O}}     \def\bO{{\bf O}}  \def\mO{{\mathscr O}}
\def\n{\nu}     \def\cP{{\mathcal P}}     \def\bP{{\bf P}}  \def\mP{{\mathscr P}}
\def\r{\varrho} \def\cQ{{\mathcal Q}}     \def\bQ{{\bf Q}}  \def\mQ{{\mathscr Q}}
\def\s{\sigma}  \def\cR{{\mathcal R}}     \def\bR{{\bf R}}  \def\mR{{\mathscr R}}
\def\S{\Sigma}  \def\cS{{\mathcal S}}     \def\bS{{\bf S}}  \def\mS{{\mathscr S}}
\def\t{\tau}    \def\cT{{\mathcal T}}     \def\bT{{\bf T}}  \def\mT{{\mathscr T}}
\def\f{\phi}    \def\cU{{\mathcal U}}     \def\bU{{\bf U}}  \def\mU{{\mathscr U}}
\def\F{\Phi}    \def\cV{{\mathcal V}}     \def\bV{{\bf V}}  \def\mV{{\mathscr V}}
\def\P{\Psi}    \def\cW{{\mathcal W}}     \def\bW{{\bf W}}  \def\mW{{\mathscr W}}
\def\o{\omega}  \def\cX{{\mathcal X}}     \def\bX{{\bf X}}  \def\mX{{\mathscr X}}
\def\x{\xi}     \def\cY{{\mathcal Y}}     \def\bY{{\bf Y}}  \def\mY{{\mathscr Y}}
\def\X{\Xi}     \def\cZ{{\mathcal Z}}     \def\bZ{{\bf Z}}  \def\mZ{{\mathscr Z}}
\def\O{\Omega}

\newcommand{\mc}{\mathscr {c}}

\newcommand{\gA}{\mathfrak{A}}          \newcommand{\ga}{\mathfrak{a}}
\newcommand{\gB}{\mathfrak{B}}          \newcommand{\gb}{\mathfrak{b}}
\newcommand{\gC}{\mathfrak{C}}          \newcommand{\gc}{\mathfrak{c}}
\newcommand{\gD}{\mathfrak{D}}          \newcommand{\gd}{\mathfrak{d}}
\newcommand{\gE}{\mathfrak{E}}
\newcommand{\gF}{\mathfrak{F}}           \newcommand{\gf}{\mathfrak{f}}
\newcommand{\gG}{\mathfrak{G}}           
\newcommand{\gH}{\mathfrak{H}}           \newcommand{\gh}{\mathfrak{h}}
\newcommand{\gI}{\mathfrak{I}}           \newcommand{\gi}{\mathfrak{i}}
\newcommand{\gJ}{\mathfrak{J}}           \newcommand{\gj}{\mathfrak{j}}
\newcommand{\gK}{\mathfrak{K}}            \newcommand{\gk}{\mathfrak{k}}
\newcommand{\gL}{\mathfrak{L}}            \newcommand{\gl}{\mathfrak{l}}
\newcommand{\gM}{\mathfrak{M}}            \newcommand{\gm}{\mathfrak{m}}
\newcommand{\gN}{\mathfrak{N}}            \newcommand{\gn}{\mathfrak{n}}
\newcommand{\gO}{\mathfrak{O}}
\newcommand{\gP}{\mathfrak{P}}             \newcommand{\gp}{\mathfrak{p}}
\newcommand{\gQ}{\mathfrak{Q}}             \newcommand{\gq}{\mathfrak{q}}
\newcommand{\gR}{\mathfrak{R}}             \newcommand{\gr}{\mathfrak{r}}
\newcommand{\gS}{\mathfrak{S}}              \newcommand{\gs}{\mathfrak{s}}
\newcommand{\gT}{\mathfrak{T}}             \newcommand{\gt}{\mathfrak{t}}
\newcommand{\gU}{\mathfrak{U}}             \newcommand{\gu}{\mathfrak{u}}
\newcommand{\gV}{\mathfrak{V}}             \newcommand{\gv}{\mathfrak{v}}
\newcommand{\gW}{\mathfrak{W}}             \newcommand{\gw}{\mathfrak{w}}
\newcommand{\gX}{\mathfrak{X}}               \newcommand{\gx}{\mathfrak{x}}
\newcommand{\gY}{\mathfrak{Y}}              \newcommand{\gy}{\mathfrak{y}}
\newcommand{\gZ}{\mathfrak{Z}}             \newcommand{\gz}{\mathfrak{z}}

\def\ve{\varepsilon}   \def\vt{\vartheta}    \def\vp{\varphi}    \def\vk{\varkappa}

\def\A{{\mathbb A}} \def\B{{\mathbb B}} \def\C{{\mathbb C}}
\def\dD{{\mathbb D}} \def\E{{\mathbb E}} \def\dF{{\mathbb F}} \def\dG{{\mathbb G}} \def\H{{\mathbb H}}\def\I{{\mathbb I}} \def\J{{\mathbb J}} \def\K{{\mathbb K}} \def\dL{{\mathbb L}}\def\M{{\mathbb M}} \def\N{{\mathbb N}} \def\O{{\mathbb O}} \def\dP{{\mathbb P}} \def\R{{\mathbb R}} \def\dQ{{\mathbb Q}}
\def\S{{\mathbb S}} \def\T{{\mathbb T}} \def\U{{\mathbb U}} \def\V{{\mathbb V}}\def\W{{\mathbb W}} \def\X{{\mathbb X}} \def\Y{{\mathbb Y}} \def\Z{{\mathbb Z}}

\newcommand{\1}{\mathbbm 1}
\newcommand{\dd}    {\, \mathrm d}



\def\la{\leftarrow}              \def\ra{\rightarrow}            \def\Ra{\Rightarrow}
\def\ua{\uparrow}                \def\da{\downarrow}
\def\lra{\leftrightarrow}        \def\Lra{\Leftrightarrow}


\def\lt{\biggl}                  \def\rt{\biggr}
\def\ol{\overline}               \def\wt{\widetilde}
\def\no{\noindent}


\let\ge\geqslant                 \let\le\leqslant
\def\lan{\langle}                \def\ran{\rangle}
\def\/{\over}                    \def\iy{\infty}
\def\sm{\setminus}               \def\es{\emptyset}
\def\ss{\subset}                 \def\ts{\times}
\def\pa{\partial}                \def\os{\oplus}
\def\om{\ominus}                 \def\ev{\equiv}
\def\iint{\int\!\!\!\int}        \def\iintt{\mathop{\int\!\!\int\!\!\dots\!\!\int}\limits}
\def\el2{\ell^{\,2}}             \def\1{1\!\!1}
\def\sh{\sharp}
\def\wh{\widehat}

\def\all{\mathop{\mathrm{all}}\nolimits}
\def\where{\mathop{\mathrm{where}}\nolimits}
\def\as{\mathop{\mathrm{as}}\nolimits}
\def\Area{\mathop{\mathrm{Area}}\nolimits}
\def\arg{\mathop{\mathrm{arg}}\nolimits}
\def\const{\mathop{\mathrm{const}}\nolimits}
\def\det{\mathop{\mathrm{det}}\nolimits}
\def\diag{\mathop{\mathrm{diag}}\nolimits}
\def\diam{\mathop{\mathrm{diam}}\nolimits}
\def\dim{\mathop{\mathrm{dim}}\nolimits}
\def\dist{\mathop{\mathrm{dist}}\nolimits}
\def\Im{\mathop{\mathrm{Im}}\nolimits}
\def\Iso{\mathop{\mathrm{Iso}}\nolimits}
\def\Ker{\mathop{\mathrm{Ker}}\nolimits}
\def\Lip{\mathop{\mathrm{Lip}}\nolimits}
\def\rank{\mathop{\mathrm{rank}}\limits}
\def\Ran{\mathop{\mathrm{Ran}}\nolimits}
\def\Re{\mathop{\mathrm{Re}}\nolimits}
\def\Res{\mathop{\mathrm{Res}}\nolimits}
\def\res{\mathop{\mathrm{res}}\limits}
\def\sign{\mathop{\mathrm{sign}}\nolimits}
\def\span{\mathop{\mathrm{span}}\nolimits}
\def\supp{\mathop{\mathrm{supp}}\nolimits}
\def\Tr{\mathop{\mathrm{Tr}}\nolimits}
\def\BBox{\hspace{1mm}\vrule height6pt width5.5pt depth0pt \hspace{6pt}}


\newcommand\nh[2]{\widehat{#1}\vphantom{#1}^{(#2)}}
\def\dia{\diamond}

\def\Oplus{\bigoplus\nolimits}




\def\qqq{\qquad}
\def\qq{\quad}
\let\ge\geqslant
\let\le\leqslant
\let\geq\geqslant
\let\leq\leqslant
\newcommand{\ca}{\begin{cases}}
\newcommand{\ac}{\end{cases}}
\newcommand{\ma}{\begin{pmatrix}}
\newcommand{\am}{\end{pmatrix}}
\renewcommand{\[}{\begin{equation}}
\renewcommand{\]}{\end{equation}}
\def\bu{\bullet}

\title[{Inverse spectral theory for the surface of revolution }]
{Inverse spectral theory and the Minkowski problem for the surface of revolution}

\date{\today}

\author[Hiroshi Isozaki]{Hiroshi Isozaki}
\address{Institute of Mathematics,
University of Tsukuba,
Tsukuba, 305-8571, Japan\\
\ isozakih@math.tsukuba.ac.jp}
\author[Evgeny L. Korotyaev]{Evgeny L. Korotyaev}
\address{
Saint-Petersburg State University, Universitetskaya nab. 7/9, St.
Petersburg, 199034, Russia, \ korotyaev@gmail.com, \
e.korotyaev@spbu.ru,}

\subjclass{}
\keywords{rotationally symmetric manifolds, inverse problem }

\begin{abstract}
\no We solve the inverse spectral problem for rotationally symmetric manifolds, which include the class of surfaces of revolution, by giving an analytic isomorphism from the space of spectral data onto the space of functions describing the radius of rotation. An analogue of the Minkowski problem is also solved.
\end{abstract}

\maketitle

\section {Introduction and main results}
\setcounter{equation}{0}


\subsection{The surface of revolution}
Suppose we are given a surface of revolution $M$  in ${\R}^{m+2}$
 with $m \geq 1$. Using the coordinates $(x,y) \in {\R}^{m+2} = {\R}^1\times{\R}^{m+1}$, $M$ is represented as
\begin{equation}
y = f(x)\omega, \quad \omega \in \S^{m}, \quad x\in I = [0,x_0],
\end{equation}
where $f \in C^2(I)$, $f(x) > 0$.
Then the induced metric on $M$ is
\begin{equation}
ds^2 =\left(1 + f'(x)^2\right)(dx)^2 + f(x)^2g_{\S^{m}},
\end{equation}
$g_{\S^{m}}$ being the standard metric on $\S^{m}$. Making the change
of variable $t = t(x)$ by
\begin{equation}
{dt\/dx} = \sqrt{1 + f'(x)^2},
\label{S1dtdx}
\end{equation}
we can rewrite $ds^2$ as
\begin{equation}
\left\{
\begin{split}
& ds^2 = (dt)^2 + r(t)^2g_{\S^{m}}, \quad r(t) = f(x(t)), \\
& 0 \leq t
\leq t_0 = \int_0^{x_0}\sqrt{1+f'(x)^2}dx.
\end{split}
\right.
\label{S1Metricrewritten}
\end{equation}
Then we have
$$
|r'(t)| < 1,
$$
since
\begin{equation}
r'(t) = f'(x(t))\frac{dx}{dt} = \frac{f'(x(t))}{\sqrt{1+f'(x(t))^2}}.
\label{S1r'(t)}
\end{equation}
Now, the Laplace-Beltrami operator on $M$ is written as
\begin{equation}
\Delta_M = \frac{1}{r^m}\partial_t\left(r^m\partial_t\right) + \frac{\Delta_Y}{r^2},
\label{S1DeltaMsurfrev}
\end{equation}
where $\Delta_Y$ is the Laplace-Beltrami operator on $\S^m$.
By imposing suitable boundary conditions on $t=0$ and $t=t_0$, one can get the spectral data for $M$. We are interested in the inverse spectral problem, i.e. the recovery of $M$ from its spectral data. Note that in this setting, we are given the operator $\Delta_Y$. The value of $t_0$ is not known a-priori, since it is computed from (\ref{S1Metricrewritten}), which contains unknown $f(x)$.
However, the eigenvalue problem for (\ref{S1DeltaMsurfrev}) is reduced to the 1-dimensional Sturm-Liouville problem, and one can  derive the value of $t_0$ from the asymptotics of eigenvalues. By virtue of (\ref{S1r'(t)}), (\ref{S1dtdx}) is rewritten as
\begin{equation}
\frac{dx}{dt} = \sqrt{1 - r'(t)^2},
\label{S1dxdt}
\end{equation}
from which one can compute $x(t)$ as well as $x_0$. We can then recover $f(x)$ from the formula $r(t) = f(x(t))$ and the inverse function theorem.

 We have thus seen that our problem is reduced to the inverse spectral problem for (\ref{S1DeltaMsurfrev}) defined on $[0,t_0]\times Y$ with suitable boundary condition. Since $t_0$ is known from the spectral asymptotics, we can assume without loss of generality that $t_0=1$. More precisely, in the general case, we have only to repeat the arguments below with
$M = [0,1]\times Y$ replaced by $M = [0,x_0]\times Y$, where $x_0$ is computed from (\ref{S1dxdt}) and $t_0$.


\subsection {Rotationally symmetric manifold}
Let us slightly generalize our problem. Assume that we are given a compact $m$-dimensional Riemannian manifold  $(Y, g_0)$
(with or without boundary). We consider  a cylindrical manifold $M = [0,1]\ts Y$  with warped product metric
\[
\lb{1}
g = (dx)^2 + r^2(x)g_0.
\]
The Laplace-Beltrami operator on $M$ is written as
\begin{equation}
\Delta_{M} = \frac{1}{r(x)^m}\partial_x\Big(r(x)^m\partial_x\Big)
+ \frac{1}{r^2(x)}\Delta_Y.
\label{S0DeltaMdefine}
\end{equation}

Two examples are given in Fig. 1, where $Y = \S^1$ and Fig. 2, where $Y = [0,\alpha]$ with a suitable boundary condition on $\partial Y$.

\begin{figure}[hbtp]
\centering
\unitlength 0.7mm 
\linethickness{0.4pt}
\ifx\plotpoint\undefined\newsavebox{\plotpoint}\fi 
\begin{picture}(75.55,65)(0,0)
\qbezier(2.4,27.25)(3,42.375)(4.8,43)
\qbezier(35.6,27.25)(36.2,42.375)(38,43)
\qbezier(61.6,27.25)(62.2,42.375)(64,43)
\qbezier(2.4,26.75)(3,11.625)(4.8,11)
\qbezier(35.6,26.75)(36.2,11.625)(38,11)
\qbezier(61.6,26.75)(62.2,11.625)(64,11)
\qbezier[30](7.5,27.25)(6.9,42.375)(5.1,43)
\qbezier[30](40.7,27.25)(40.1,42.375)(38.3,43)
\qbezier(66.7,27.25)(66.1,42.375)(64.3,43)
\qbezier[30](7.5,26.75)(6.9,11.625)(5.1,11)
\qbezier[30](40.7,26.75)(40.1,11.625)(38.3,11)
\qbezier(66.7,26.75)(66.1,11.625)(64.3,11)
\qbezier(4.65,43)(21.075,58.625)(39,42.75)
\qbezier(4.65,11)(21.075,-4.625)(39,11.25)
\put(74.5,24){\makebox(0,0)[cc]{$x$}}
\put(1.25,22.25){\makebox(0,0)[cc]{$0$}}
\put(25,55){\makebox(0,0)[cc]{$r(x)$}}
\qbezier(39,42.75)(49.5,34.25)(64,42.75)
\qbezier(39,11.25)(49.5,19.75)(64,11.25)
\put(64.25,24.75){\makebox(0,0)[cc]{$1$}} \linethickness{0.2pt}
\put(5.25,64.7){\line(0,-1){60.4}} \put(3.15,27.5){\line(1,0){72.4}}
\multiput(7.75,29.25)(-.0517578125,-.0336914063){512}{\line(-1,0){.0517578125}}
\end{picture}
\caption{\footnotesize The surface with $Y=\{y\in \S^1\}$.}
\lb{F1}
\end{figure}
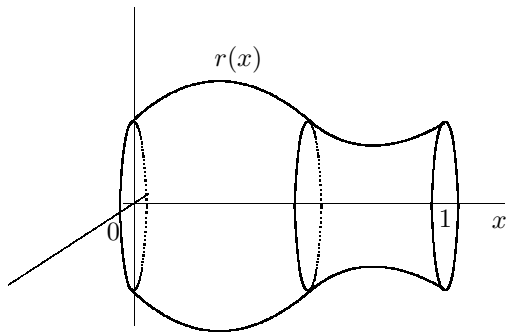

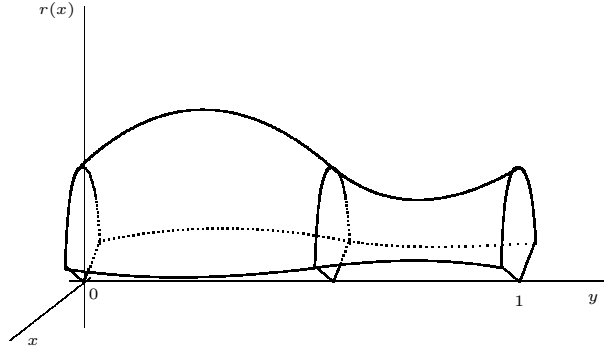
\begin{figure}[hbtp]
\tiny
\unitlength 0.7mm 
\ifx\plotpoint\undefined\newsavebox{\plotpoint}\fi 
\begin{picture}(115.77,67.275)(0,0)
\linethickness{0.2pt} \put(14.41,14.775){\line(1,0){101.36}}
\put(17.35,6.025){\line(0,1){60.9}}
\multiput(18.4,15.3)(-.0431303116,-.0337110482){353}{\line(-1,0){.0431303116}}
\put(19.1,12.325){\makebox(0,0)[cc]{$0$}}
\put(12.1,66.275){\makebox(0,0)[cc]{$r(x)$}}
\put(99.95,10.925){\makebox(0,0)[cc]{$1$}}
\put(7.6,3.275){\makebox(0,0)[cc]{$x$}}
\put(113.95,11.4){\makebox(0,0)[cc]{$y$}} \linethickness{0.6pt}
\qbezier(16.51,36.475)(39.505,58.35)(64.6,36.125)
\qbezier(64.6,36.125)(79.3,24.225)(99.6,36.125)
\multiput(99.95,14.425)(.033510638,.0875){94}{\line(0,1){.0875}}
\qbezier[5](66,18)(66.7,20)(67.4,22)
\multiput(64.6,14.425)(.033510638,.0875){44}{\line(0,1){.0875}}
\multiput(17.175,14.425)(.033510638,.0875){30}{\line(0,1){.0875}}
\qbezier[5](18.5,17.5)(19.2,19.5)(19.9,21.5)
\multiput(99.95,14.6)(-.039095745,.033510638){94}{\line(-1,0){.039095745}}
\multiput(64.6,14.6)(-.039095745,.033510638){94}{\line(-1,0){.039095745}}
\multiput(17.175,14.6)(-.039095745,.033510638){94}{\line(-1,0){.039095745}}
\qbezier(103.1,21.775)(102.313,36.213)(99.775,36.3)
\qbezier[30](67.75,21.775)(66.963,36.213)(64.425,36.3)
\qbezier[30](20.325,21.775)(19.537,36.213)(17,36.3)
\qbezier(99.6,36.3)(96.975,36.388)(96.45,17.225)
\qbezier(64.25,36.3)(61.625,36.388)(61.1,17.225)
\qbezier(16.825,36.3)(14.2,36.388)(13.675,17.225)
\qbezier(13.675,17.05)(34.412,13.813)(61.1,17.225)
\qbezier(60.925,17.225)(81.225,20.025)(96.625,17.225)
\qbezier[50](20.15,22.3)(43.338,27.2)(67.575,22.3)
\qbezier[30](67.75,22.3)(79.3,20.375)(102.75,21.95)
\end{picture}
\caption{\footnotesize The surface of revolution of an angle
$\a<\pi$} \lb{tune}
\end{figure}

\noindent
For the operator (\ref{S0DeltaMdefine}), we impose one of the following boundary conditions on $\partial M = \{0,1\}\times Y$ : For $y \in Y$,
\[
\label{S1BC}
\left\{
\begin{array}{l}
{\rm Dirichlet \ b. c. } \quad f(0,y) = f(1,y) = 0,\\
{\rm Mixed\ b. c.} \quad f(0,y)=0, \ f'(1,y)+b f(1,y)=0,  \  b\in \R, \\
 {\rm Robin \ b. c.} \quad f'(0,y)-a f(0,y)=0, \
 f'(1,y)+b f(1,y)=0, \  a,b\in \R.
\end{array}
\right.
\]

The Laplacian $-\D_Y$ on $Y$ has the discrete spectrum
$$
0\le
E_1\le E_2\le E_3\le . . .
$$
 with an associated orthonormal family of
eigenfunctions ${\P_\n, \n\ge 1}$, in $L^2(Y )$. Then, we have the orthogonal decomposition
$$
L^2(M)=\os_{\n\ge
1} \mL_\n^2(M),
$$
$$
\mL_\n^2(M)=\rt\{h(x,y)=f(x)\P_\n(y)\, ;\,  \int_0^1|f(x)|^2r^m(x)dx < \infty
\rt\},\qq \n\ge 1.
$$
Thus, $-\Delta_M$
 is unitarily equivalent to a
direct sum of one-dimensional  operators,
$$
\lb{3} -\D_M\backsimeq\os_{\n=1}^{\iy}\left(-\D_\n\right),
$$
\begin{equation}
\label{S1-Deltanu1}
\begin{aligned}
 -\D_\n =-{1\/\r^2} \pa_x \left(\r^2 \pa_x\right) +{E_\n\/r^2} ,\quad
\r=r^{m/2}, \quad {\rm on} \quad L^2\big([0,1];r^m(x)dx\big).
\end{aligned}
\end{equation}
We call $-\Delta_{\nu}$ a Sturm-Liouville operator.
The boundary condition (\ref{S1BC}) is inherited for $-\Delta_{\nu}$ :
\[
\lb{bc123}
\left\{
\begin{aligned}
& {\rm Dirichlet \ b. c. } \quad f(0)=f(1)=0,\\
&{\rm Mixed \ b. c.} \quad f(0)=0,\ f'(1)+b f(1)=0,\ b\in \R,\\
&{\rm Robin \ b. c.}\quad f'(0)-af(0)=0,\ f'(1)+b f(1)=0, \ a,b\in\R.
\end{aligned}
\right.
\]

The  operator $-\D_\n$ actually depends on
\[
{\r'\/\r} \quad {\rm and} \quad \rho(0) = r(0)^{m/2}.
\]
For our purpose, it is convenient to introduce a parameter $q_0 = \rho'(0)/\rho(0)$ and put
\begin{equation}
\frac{\rho'(x)}{\rho(x)} = q_0 + q(x).
\end{equation}
Then  $r(x)$ is written as
\[
\lb{2}
 r(x)=r(0)e^{2Q(x)/m},\quad Q(x)=\int_0^x(q_0+q(t))dt.
\]
We then have
\begin{equation}
\frac{r'(0)}{r(0)}  = \frac{2q_0}{m}, \quad \quad
\log\frac{r(1)}{r(0)} = \frac{2}{m}\Big(q_0 + \int_0^1q(x)dx\Big).
\label{S1r'(0)r(1)}
\end{equation}
This implies that, if we are given either $r(0)$ and $r'(0)$, or $r(0)$ and $r(1)$, we can reconstruct $r(x)$ from  $q(x)$ for $0 \leq x \leq 1$.

The problem we address in this paper is the characterization of the range of  the {\it spectral data mapping}
$$
q \to \left\{\mu_n(q), \kappa_n(q)\right\}_{n=1}^{\infty},
$$
where $\mu_n$ and $\kappa_n$ are eigenvalues and norming constants for
(\ref{S1-Deltanu1}) with a fixed $\nu$.

\subsection{Function spaces}
Let us introduce the following spaces of real functions
\begin{equation}
\begin{aligned}
\lb{dWH}
&\mW_1^0 =\rt\{q \in L^2(0,1)\ ;\ q'\in L^2(0,1), \ q(0)=q(1)=0\rt\}, \qqq
\\
&\mH_\alpha =\rt\{q\in L^2(0,1)\ ; \  q^{(\a)} \in L^2(0,1), \int
_0^1q^{(j)}(x)dx=0, \forall\ j=0,..,\a\rt\},
\end{aligned}
\end{equation}
where $\alpha \geq 0$, equipped with norms
$$
\|q\|^2_{\mW_1^0}=\|q'\|^2=\int_0^1|q'(x)|^2dx,\qqq
\|q\|^2_{\mH_\a}=\|q^{(\a)}\|^2=\int_0^1|q^{(\a)}(x)|^2dx.
$$
Define the spaces of even functions $L_{even}^2(0,1)$,  and of odd functions
$L_{odd}^2(0,1)$  by
\[
\begin{aligned}
\lb{oeL}
L_{even}^2(0,1)&=\rt\{q\in L^2(0,1)\, ; \, q(x)=q(1-x), \qq \forall
\ x\in (0,1)\rt\},\\
L_{odd}^2(0,1)&=\rt\{q\in L^2(0,1)\, ; \, q(x)=-q(1-x), \qq \forall
\ x\in (0,1)\rt\},\\
L^2(0,1)&=L_{even}^2(0,1) \os  L_{odd}^2(0,1)
\end{aligned}
\]
and for $\o=even$ or  $\o=odd$ we define
\[
\lb{oe}
\mW_1^{0,\o}= \mW_1^0 \cap L_{\o}^2(0,1),\quad
\mH_\a^{\o} =\mH_\a\cap L_{\o}^2(0,1), \qq \a\ge 0.
\]

We also introduce the space $\ell^2_{\a}$ of real sequences
$h=(h_n)_1^{\iy }$,
equipped with the norm
\[
\lb{1.0}
\|h\|_{\a}^2=2\sum _{n\ge 1}(2\pi n)^{2\a}|h_n|^2,\qqq  \a\in \R,
\]
and  let $\ell^2=\ell_0^2$.  Finally we define the set $\cM_1 \ss\ell^2$  by
\[
\lb{S1Mj} \cM_1 =
\cM_1\left((\mu_n^0)_{n=1}^{\infty}\right)=\left\{(h_n)_{n=1}^{\iy}\in\ell^2\,
; \,\m_1^0\!+\! h_1\!<\!\m_{2}^0\!+\! h_{2}\! <\dots \right\},
\]
where the sequence $(\mu_n^0)_{n=1}^{\infty}$ will be specified below.

\subsection{Main results I. Spectral data mapping}
We are now in a position to stating our main results of this paper.

\subsubsection{ Dirichlet boundary condition}
First we consider   $-\D_\n, \n\ge 1$, on the
interval $[0,1]$  with  Dirichlet boundary condition:
\[
\lb{ipC11}
\left\{
\begin{split}
&-\D_\n f=-{1\/\r^2}(\r^2f')'+{E_\n\/r^2}f, \quad \r=r^{m/2}, \quad {\rm on} \quad (0,1),\\ &f(0)=f(1)=0.
\end{split}
\right.
\]
Denote by $\m_n=\m_n(q), n=1,2\cdots$,
the eigenvalues of $-\D_{\nu}$.  It is well-known that all $\m_n$ are
simple and satisfy
\begin{equation}
\lb{ipD12}
\begin{split}
&\m_n=\m_n^0+c_0+\wt\m_n, \\
& \m_n^0 = (n\pi)^2, \quad (\wt\m_n)_{1}^{\iy}\in\ell^2, \\ &c_0=\int_0^1\Big((q_0+q)^2+{E_\n\/r^2}\Big)dx,
\end{split}
\end{equation}
where $\m_n^0$, $n\ge 1$, are the eigenvalues for the  unperturbed case $r=1$.
Following \cite{PT87}, \cite{IK13}, we introduce the norming constants
\[
\lb{ipC13} \vk_n(q)=\log\left|\r(1)f_n'(1,q)\/f_n'(0,q)\right|,
\qquad n\ge 1,
\]
where $f_n$ is the $n$-th eigenfunction of $-\Delta_{\nu}$. Note that $f_n'(0)\ne 0$ and
 $f_n'(1)\ne 0$. Recall that
\begin{equation}
q_0 = \frac{\rho'(0)}{\rho(0)}.
\label{Defineq0}
\end{equation}

\begin{theorem}
\label{T1}
  Fix $\n\ge 1$, and consider
$-\D_\n$ with the Dirichlet boundary condition.
 Assume either (i) or (ii) of the following conditions:

\noindent
(i) $q_0=0$,

\noindent
(ii) $\n=1$ and $E_1=0$.

Then the  mapping
$$
\P : q\mapsto \Big((\wt\m_{n}(q))_{n=1}^{\iy}\,,
(\vk_{n}(q))_{n=1}^{\iy}\Big)
$$
defined by \er{ipD12}, \er{ipC13} is a real-analytic isomorphism
between $\mW_1^0$ and $\cM_1\ts \el2_1$, where $\mathcal M_1$ is
defined by (\ref{S1Mj}) with $\m_n^0=(\pi n)^2,
n\ge 1$. In particular, in the symmetric case (the function $q$ is
odd and the manifold $M$ is symmetric with respect to the plane
$x={1\/2}$) the spectral data mapping
\[
\wt\m: \mW_1^{0,odd} \ni q \to (\widetilde\mu_n)_1^\iy\in \cM_1
\]
is a  real analytic isomorphism between $\mW_1^{0,odd}$ and  $\cM_1$.
\end{theorem}

\subsubsection{Mixed boundary condition}
We next consider  $-\D_\n,
\n\ge 1$, with  mixed boundary condition:
\[
\lb{ipC21}
\left\{
\begin{aligned}
&- \Delta_{\nu}f = -{1\/\rho^2}(\rho^2f')'+{E_\n\/r^2},\quad \r=r^{m/2} \quad {\rm on} \quad (0,1),\\
& f(0)=0,\quad f'(1)+b
f(1)=0, \quad
(b,q)\in \R\ts\mW_1^0.
\end{aligned}
\right.
\]
 Let $\mu_n=\mu_n(q,b), n=0,1,2,...$ be the associated eigenvalues. They satisfy
\[
\lb{ipC22}
\begin{split}
& \mu_n(q,b)=\mu_n^0+c_0+\wt\mu_n(q,b),\\
&\mu_n^0 = \pi^2(n+{1\/2})^2+2b, \\
&(\wt\mu_n)_{1}^{\iy}\in\ell^2,\qq c_0=\int_0^1\rt((q_0+q)^2+{E_\n\/r^2}\rt)dx.
\end{split}
\]
where $\mu_n^0$'s are the eigenvalues for  for the unperturbed case $r=1$.
As in \cite{KC09}, we introduce the norming constants
\[
\label{ipC23}
\c_n(q,b)=\log\left|\r(1)f_n(1,q,b)\/f_n'(0,q,b)\right|,\qquad n\ge
0,
\]
where $f_n$ is the $n$-th  eigenfunction satisfying $f_n'(0,q,b)\ne 0$
 and $f_n(1,q,b)\ne 0$.  When $q =b = 0$, a simple calculation gives
\[
\label{ipC24}
\c_n^0:=\c_n(0,0)=-\log \pi(n\!+\!{\textstyle{1\/2}}).
\]

\begin{theorem}
\lb{T2}
 For any fixed $(b,q_0,\n)\in \R^2\ts\N$, consider
$-\D_\n$ with mixed boundary condition. Assume either (i) or (ii)  of the following conditions:

\noindent
(i) $q_0=0$,

\noindent
(ii) $\n=1$ and $E_1=0$.

Then the mapping defined by \er{ipC22}-\er{ipC24}
$$
\P:q\mapsto
\left((\wt\mu_n(q,b))_{n=1}^{\iy}\,,(\c_{n-1}(q,b)-\c_{n-1}^0)_{n=1}^{\iy}\right)
$$
is a real-analytic isomorphism between $\mW_1^0$ and $\mathcal M_1\ts\ell^2_1$, where
$\mathcal M_1$ is defined by (\ref{S1Mj})
 with $\m_n^0=(\pi n+{1\/2})^2+2b,
n\ge 1$. Moreover, for each $(q,b)\in \mW_1^0\ts\R$, the following
identity holds:
\[
\label{S1IdentityB}
b=\sum_{n=0}^{\iy} \lt(2-{e^{\c_n(q,b)}\/|{\pa w\/\pa \l}(\mu_n,q,b)|}\rt),
\]
where the function $w(\l,q,b)$ is given by
\[
\label{S1adam} w(\l,q,b)=
\cos\sqrt\l\cdot\prod_{n=0}^{\iy}{\l-\mu_n(q,b)\/\l-\mu_n^0}\,,\qquad \l\in\C.
\]
Here  (\ref{S1IdentityB}) and  (\ref{S1adam}) converge uniformly on any bounded
subsets in ${\bf C}$.
\end{theorem}

\subsubsection{Robin boundary conditions}
The 3rd case is the  Robin  boundary condition:
\[
\lb{ipC31}
\left\{
\begin{aligned}
&- \Delta_{\nu}f = -{1\/\rho^2}(\rho^2f')'+{E_\n\/r^2},\quad \r=r^{m/2} \quad {\rm on} \quad (0,1), \\
& f'(0)-af(0)=0,\quad
f'(1)+b f(1)=0,\quad (a,b,q)\in \R^2\ts\mW_1^0.
\end{aligned}
\right.
\]
 Let $\m_n=\m_n(q,a,b), n=0,1,2,...$ be the associated eigenvalues.
  It is well-known that
\[
\lb{Case3EV}
\begin{split}
& \m_n= \mu_n^0+c_0+\wt\m_n(q,a,b), \\
&\m_n^0 = (n\pi)^2+2(a+b), \quad (\wt\m_n)_{1}^{\iy}\in\ell^2, \\ & c_0=\int_0^1\rt((q_0+q)^2+{E_\n\/r^2}\rt)dx.
\end{split}
\]
Note $\m_n^0, n\ge 0$   are  the eigenvalues  for $r=1$.
The norming constants are defined by
\[
\label{Case3NC}
\f_n(q,a,b)=\log\left|\r(1)f_n(1,q,a,b)\/f_n(0,q,a,b)\right|,\qquad
n\ge 0,
\]
where $f_n$ is the $n$-th eigenfunction. They satisfy
$f_n(1,q,a,b)\ne 0$ and $f_n(0,q,a,b)\ne 0$.

\begin{theorem}
\label{T3}
For any fixed $(a,b,q_0,\n)\in \R^3\ts\N$,
consider $-\D_\n$ with Robin boundary condition.
Suppose either (i) or (ii) of the following conditions hold:

\noindent
(i) $q_0=0$,

\noindent
(ii) $\n=1$ and $E_1=0$.

Then the mapping defined by \er{Case3EV}, \er{Case3NC}
\[
\lb{IPGbct}
 \P_{a,b}:q\mapsto \Big((\wt\m_{n}(q,a,b))_{n=1}^{\iy}\, ,
(\f_{n}(q,a,b))_{n=1}^{\iy}\Big)
\]
is a real-analytic isomorphism between $\mW_1^0$ and $\cM_1\ts
\el2_1$, where $\mathcal M_1$ is defined by (\ref{S1Mj})  with $\m_n^0=(\pi n)^2+2(a+b), n\ge 1$.
\end{theorem}
\no {\bf Remark.} 1) In Theorems \ref{T1}-\ref{T3}  we consider two
cases: (i) $q_0=0$, or (ii) $\n=1$ and $E_1=0$. The inverse problems
for the cases: (1) $q_0\in \R, \n\ge 2$ or (2) $\n=1$ and $E_1\ne 0$
in Theorems \ref{T1}-\ref{T3} are still open.

2) We have the standard asymptotics \er{ipD12}, \er{ipC22} and
\er{ipC31} for fixed $\n$. It is interesting to determine the
asymptotics uniformly in $\n\ge 1$.


\subsection {Main results II. The curvature mapping.}

The Minkowski problem in classical differential geometry asks the existence of a convex surface with a prescribed Gaussian curvature. More precisely, for a given strictly positive real function $F$ defined on a sphere, one seeks a strictly convex compact surface
$\cS$, whose Gaussian curvature at $x$ is equal to
$F({\bf n}(x))$, where ${\bf n}(x)$ denotes the outer unit normal to
$\cS$ at $x$. The Minkowski problem was solved by Pogorelov \cite{P74} and
 by Cheng-Yau \cite{CY76}.

We consider only the case $m=\dim Y=1$.
Note that our surface is not convex, in general.
 We solve an analogue of the Minkowski problem
in the case of the surface of revolution by showing the existence of
a bijection between the  Gaussian curvatures and the profiles of surfaces.

As it is well-known, the Gaussian  curvature $\cK$ is given by
\[
\lb{K1} \cK=-{r''\/r},\qqq \r=r^{1\/2}.
\]
As above, we represent the profile $r(x)$ in the following way:
\[
\lb{K1r}
 r(x)=r_0e^{2Q(x)}, \quad Q(x)=\int_0^x (q_0+q(t))dt,\quad (q_0,q)\in \R\ts \mW_1^0.
\]
 Then we have
\[
\lb{K2}
\cK=-2q'-4(q_0+q)^2.
\]
Note that if $q=0$, then $\cK=-4q_0^2<0$ is a negative constant.
Letting
\[
\lb{K3}
 \cK_0=4\int_0^1
 (2q_0q+q^2)dx,\quad
G(q)=2q'+4(q_0+q)^2-\cK_0,
\]
we  rewrite $\cK$ into the form
\begin{equation}
\lb{K4}
\cK=-G(q)-\cK_0-4q_0^2.
\end{equation}


\begin{theorem}
\lb{T4} Let
the Gaussian  curvature $\cK$ and the profile $r(x)$ be given by
\er{K1},
 \er{K1r}, where $(q_0,q)\in \R\ts\mW_1^0$.
Then the mapping $G: \mW_1^0\to \mH_0$ defined by
\[
q\to G(q)=-\cK-\cK_0-4q_0^2
\]
is a real  analytic isomorphism between
$\mW_1^0$ and  $\mH_0$. Moreover, the constant $\cK_0$ is uniquely
defined by $G(q)$.
\end{theorem}

\noindent
{\bf Remark.} This theorem also holds with $\mW_1^0$ replaced by $\mH_1$.

\medskip

Theorem \ref{T4} gives the mapping between $\cK$ and the profile $r$.
Thus Theorems \ref{T1} $\sim$ \ref{T4} make  the mapping
$$
 {\rm Gaussian \ curvature} \ \cK\qq \to \qq {\rm eigenvalues\ +\  norming
 \ constants}
$$
well-defined. We illustrate this by Theorem \ref{T5}.
We consider the Sturm-Liouville problem with Robin  boundary condition:
\[
\lb{ipGc1}
-{1\/\r^2}(\r^2f')'+{E_\n\/r^2}f=\l f,\qq f'(0)-af(0)=0,\quad f'(1)+b f(1)=0.
\]
Let $\x = G(q) \in \mH_0$ and $A=(a,b,q_0)\in\R^3$. Let
$\m_n=\m_n(\x,A), n=0,1,2,...$ be the eigenvalues of \er{ipGc1}.
They satisfy
\begin{equation}
\lb{ipCc2}
\begin{split}
& \m_n(\x,A)= \mu_n^0+c_0+\wt\m_n(\x,A), \\
& \mu_n^0 = (n\pi)^2 + 2(a+b), \quad
 (\wt\m_n)_{1}^{\iy}\in\ell^2,\\
&c_0=\int_0^1\rt(q^2+{E_\n\/r^2}\rt)dx.
\end{split}
\end{equation}
Here  $\mu_n^0$, $n\ge 0$, are the unperturbed eigenvalues for the
case $r=1$. We introduce the norming constants
\[
\label{ipGc3}
\f_n(\x,A)=\log\lt|{\r(1)f_n(1,\x,A)\/f_n(0,\x,A)}\rt|,\qq n\ge 1,
\]
where $f_n$ is the $n$-th eigenfunction. Note that  $f_n(1,\x,A)\ne
0$ and $f_n(0,\x,A)\ne 0$.


\begin{theorem}
\label{T5}
Let $A=(a,b, q_0)\in\R^3, \n\ge 1$ be fixed and  consider $-\D_\n$ with Robin boundary condition.   Assume either (i) or (ii) of the following conditions:

\noindent
(i) $q_0=0$,

\noindent
(ii) \ $\n=1$ and $E_1=0$.

Then the mapping defined by \er{ipCc2}, \er{ipGc3}
\[
\lb{ipCc4} \x\to \F_{A}(\x)=\Big((\wt\m_{n}(\x,A))_{n=1}^{\iy}\, ,
(\f_{n}(\x,A))_{n=1}^{\iy}\Big)
\]
is a real-analytic isomorphism between $\mH_0$ and $\cM_1\ts
\el2_1$, where $\mathcal M_1$ is defined by  (\ref{S1Mj})
 with $\m_n^0=(\pi n)^2+2(a+b), n\ge 1$.
\end{theorem}

\subsection{Brief overview}
There is an abundance of works devoted to the spectral theory and inverse problems for the surface of revolution from the view points of classical inverse Strum-Liouville theory, integrable systems, micro-local analysis, see \cite{AA07}, \cite{E98}, \cite{GWL05}, \cite{GL99}, \cite{M83} and references therein.
Bruning-Heintz \cite{BH84} proved that the symmetric metric is determined from the spectrum by using the 1-dimensional Gel'fand-Levitan theory \cite{L84}, \cite{M77}. 

For integrable systems associated with
surfaces of revolution, see e.g. \cite{KT96}, \cite{Ta97},\cite{BeKo99},
\ \cite{SW03}, \cite{S08} and references therein. Here we mention the work of Zelditch \cite{Z98}, which proved that the isospectral  revolutionary surfaces of simple length spectrum, with some additional conditions,  are isometric. In fact, the assumptions ensure the existence of global action-angle variables for the geodesic flow, which entails that the Laplacian has a global quantum normal form in terms of action operators. From the singularity expansion of the trace of wave group, one can then reconstruct the global quantum normal form, hence the metric. This argument, in due course, recovers the result of \cite{BH84}. Note, however, that the class of metrics considered is shown to be residual in
the class of metrics satisfying all the assumptions above concerning
the metric but not the simple length spectrum assumption.

In the proof we use the analytic approach of Trubowitz and his co-authors
(see \cite{PT87} and references therein) plus its development for periodic systems \cite{KK97}. Using them we obtain
the global transformation for  inverse Sturm-Liouville theory \cite{IK13}.
Note that for \cite{IK13} the results of inverse Sturm-Liouville theory \cite{PT87}, \cite{KC09} and \cite{K02} are important.

\subsection{Plan of the paper}
We start from proving Theorem \ref{T4}, which is based on
an abstract theorem in non-linear functional analysis \cite{KK97}.
In Section 2, we do it after preparing the estimates for the
 Riccati type mapping. The idea of the proof of Theorems 1.1,
  1.2 and 1.3 consists in converting the Sturm-Liouville equation
$$
- \frac{1}{\rho^2}\left(\rho^2f'\right)' + \frac{E}{r^2} = 0
$$
 to the Schr{\"o}dinger equation
$$
- y'' + py = Ey
$$
    using some non-linear mapping. In Section 3, we explain the results for the isomorphic property of the spectral data mapping.  The paper \cite{IK13} has been prepared for this purpose, and using the results there we shall prove Theorem 1.1, 1.2. and 1.3 in Section 4.

\section {The  curvature inverse problem and Riccati type mappings}
\setcounter{equation}{0}

\subsection{Estimates for Riccati type mappings.}
We define the  mapping $G: \cH\to \mH_0$, where $\cH=\mH_1$  or
$\cH=\mW_1^0$  by

\[
\lb{dR}
\begin{aligned}
\ca p=G(q)=q'+q^2+2q_0q-c_0,\qqq c_0=\int_0^1 (q^2+2q_0q)dx, \\
 q_0=\const \in \R, \qq  q\in\mW_1^0\qq or \qq q\in\mH_1    \ac,
\end{aligned}
\]


\begin{lemma}
\lb{TR1} Let $p$ be given by \er{dR}, where $ q\in \mH_1$ or $q\in
\mW_1^0$. Then the following estimates hold true:
\[
\lb{Rx1} \|q'\|^2\le  \|p\|^2=\|q'\|^2+\|q^2+2q_0q-c_0\|^2,
\]
\[
\lb{Rx2} \|p\|^2=\|q'\|^2+\|q^2\|^2+4q_0^2\|q\|^2+4q_0(q^3,1)-c_0^2,
\]
\[
\lb{Rx3} \|p\|^2\le \|q'\|^2+\|q^2\|^2+4q_0^2\|q\|^2+4q_0(q^3,1),
\]
where $(\cdot, \cdot)$ is the scalar product in  $L^2(0,1)$.
\end{lemma}
\no {\bf Proof.} Let $h=q^2+2q_0q-c_0$. We have
$$
\begin{aligned}
&\|p\|^2=\|q'\|^2+\|h\|^2+2(q',h),\\
&(q',h)=(q',q^2+2q_0q-c_0)=0,\\
 \end{aligned}
$$
where the integration by parts has been used. This yields \er{Rx1}.
We have
$$
\begin{aligned}
&\|h\|^2=\|q^2+2q_0q-c_0\|^2=\|q^2+2q_0q\|^2-2(q^2+2q_0q,c_0)+c_0^2
\\
=\|q^2+2q_0q\|^2-c_0^2
&=\|q^2\|^2+4q_0^2c_0+4q_0(q^2,q)-c_0^2,\\
& \|q^2+2q_0q\|^2=\|q^2\|^2+4q_0^2\|q\|^2+4q_0(q^2,q)=
\|q^2\|^2+4q_0^2\|q\|^2+4q_0(q^3,1)
\end{aligned}
$$
and together with \er{Rx1} this yields \er{Rx2} and \er{Rx3}.  \BBox

\medskip
We show that that mapping $G= G(q)=G(q,q_0)$ in \er{dR} is real analytic.

\begin{lemma}
\lb{TR2} Let $\cH=\mH_1$  or $\cH=\mW_1^0$ and let $q_0\in \R$. The
mapping $G: \cH\to \mH_0$ given by \er{dR} is real analytic and its
gradient is given by
\[
\lb{ar1}
 {\pa G(q)\/\pa q} f=f'+2(q_0+q) f -\int_0^1 2(q_0+q)fdx,\qqq
  \qqq \ \forall \ \ q, f\in \cH.
\]
Moreover, the operator ${\pa G(q)\/\pa q} $ is invertible for all
$q\in\cH$.
\end{lemma}

\no {\bf Proof}. By the standard arguments (see \cite{PT87}), we see
that $G(q)$ is real analytic  and its gradient is given by \er{ar1}.

 Due to \er{ar1}, the linear operator ${\pa G(q)\/\pa q}: \cH\to
\mH_0$ is a sum of a boundedly invertible operator and a
compact operator for all $q\in \cH$. Hence ${\pa G(q)\/\pa q}$  is a
Fredholm operator. We prove that the operator ${\pa G(q)\/\pa q}$ is
invertible by contradiction. Let $f\in \cH$ be a solution of the
equation
\[
\lb{ar2} {\pa G(q)\/\pa q}f=0, \qqq f\ne 0,
\]
for some fixed $q\in \cH$.  Due to \er{ar1} we have the equation
\[
\lb{ar3} {\pa G(q)\/\pa q}f=f'+2(q_0+q) f -C=0,\qqq C=\int_0^1
2(q_0+q)fdx.
\]
 This implies
\[
(e^{2Q}f)'=Ce^{2Q},\qqq Q=\int_0^x(q_0+q)dt.
\]

Let us first assume that the constant $C=0$. Then we get $ (e^{2Q}f)'=0 $, which
yields $(e^{2Q}f)(x)=(e^{2Q}f)(0), x\in [0,1]$. If $f\in \mW_1^0$,
then we obtain $(e^{2Q}f)=0$ and $f=0$. If $f\in \mH_1$, then we
obtain $(e^{2Q}f)(x)=f(0)$ and $f=e^{-2Q}f(0)$. This gives $f=0$,
since $\int_0^1fdt=0$. In any case, we have arrived at a
contradiction.

Next let us assume that $C\ne 0$. Without loss of generality, we can assume that
$C=1$. Then we get
$$
(e^{2Q}f)(x)=f(0)+\int_0^xe^{-2Q}dt.
$$
If $f\in \mW_1^0$, then we obtain
$(e^{2Q}f)(1)=\int_0^1e^{-2Q}dt>0$, which gives
a contradiction.

If $f\in \mH_1$, then we obtain
$(e^{2Q}f)(1)=f(0)+\int_0^1e^{-2Q}dt>f(0)$, which again gives
a contradiction.  Thus the operator ${\pa G\/\pa q} $ is invertible
for all $q\in\cH$.
\BBox


\subsection {Analytic isomorphism}

In order to prove  Theorem \ref{T4} we use the "direct approach"
in \cite{KK97} based on nonlinear functional analysis. Our main tool is the following theorem in \cite{KK97}.


\begin{theorem}
\lb{TA97}
 Let $H, H_1$ be real separable Hilbert spaces equipped with norms
$\|\cdot \|, \|\cdot \|_1$. Suppose that the map $f: H \to H_1$
satisfies the following conditions:

\no i) $f$ is real analytic  and the operator ${d \/dq}f$ has an
inverse for all $q\in H$,

\no ii) there is a nondecreasing function $\e: [0, \iy ) \to [0, \iy
), \e (0)=0,$ such that $\|q\|\le \e (\|f(q)\|_1)$  for all $q\in
H$,

\no iii) there exists a linear isomorphism $f_0:H\to H_1$ such that
the mapping $f-f_0: H \to H_1$  is compact.

\no Then $f$ is a real analytic isomorphism between $H$ and $H_1$.
\end{theorem}

{\bf Proof of Theorem \ref{T4}.} We check all conditions in Theorem
\ref{TA97} for the mapping $\x=G(q), q\in \mW_1^0$ given by \er{K3}.
The proof for the case $q\in \mH_1$ is similar. We rewrite this
mapping in the form
$$
\x=G(v/2)=v'+2v_0v+v^2-c_0,\qqq c_0=\int_0^1 (2v_0v+v^2)dt,
$$
where  $v=2q\in \mW_1^0$ and $v_0=2q_0$ is a constant.

 Lemma \ref{TR2} implies the assertion (i), and  Lemma \ref{TR1} the assertion (ii).
Let us  check iii). We take a model mapping $\x_0$ by $\x_0(v)=v'$.
Suppose $q^\n\to q$ weakly in $\mW_1^0$ as $\n\to \iy$. Then
$q^\n\to q$ strongly in $\mH_0$ as $\n\to \iy$, since the imbedding
mapping $\mW_1^0\to\mH_0$ is compact. Hence the mapping $q\to
\x(v)-\x_0(v)$ is compact.

Therefore, all conditions in Theorem \ref{TA97} hold true and the mapping
 $G:\mW_1^0 \to \mH_0$ is a real analytic isomorphism  between
  $\mW_1^0$ and $\mH_0$.
\BBox


\section{Spectral data mapping for the case $\nu=1$ and $E_1=0$}
\setcounter{equation}{0}

\subsection {Unitary transformations.}
Consider the Sturm-Liouville operator  $-\D_{q}$ defined in
  $L^2((0,1);\r^2dx)$, where $\r=\r(x)>0$, having the form
\[
\lb{if1}
-\D_{q} f=-{1\/ \r^2}(\r^2f')',\qqq\qqq \r=r^{m\/2}=e^Q,
\]
equipped with the boundary condition
\[
\lb{if2}
f'(0)-af(0)=0,\qquad f'(1)+b f(1)=0,\qquad a,b\in \R\cup \{\iy\}.
\]
Here $Q'$ is continuous on $[0,1]$.
 We define the simple unitary transformation $\mU$ by
\[
\lb{dUx}
 \mU:
L^2([0,1],\r^2dx)\to L^2([0,1],dx),\qqq \mU f= \r f.
\]
 We transform the operator $-\D_{q}$ into the Schr\"odinger operator $S_p$ by
\[
\lb{5}
\begin{aligned}
\mU (-\D_{q}) \mU^{-1}=
 -\r^{-1}\pa_x \r^2\pa_x \r^{-1}=\cD^*\cD=S_p+c_0,\qq
S_p =-{d^2\/dx^2}+p,\\
 c_0=\int_0^1(Q''+(Q')^2)dx,\qqq
 \qqq p=Q''+(Q')^2-c_0.
 \end{aligned}
\]
since using the identity $\r=\r_0e^{Q}$ we obtained
\[
\begin{aligned}
\lb{a7}
\cD=\r\ \pa_x \ \r^{-1}= \pa_x -Q',\qqq \cD^*=
\rt(\r \ \pa_x \ \r^{-1}\rt)^*=-\pa_x-Q',\\
\cD^*\cD=-(\pa_x +Q')(\pa_x -Q')=-\pa_x^2+Q''+(Q')^2.
\end{aligned}
\]
Here the operator $S_p=-{d^2\/dx^2}+p$ acts in $ L^2([0,1],dx)$. We
describe the boundary conditions for the operators $\D_{q}f$ and
$S_py$, where $y=\r f$. We have the following identities
\[
\begin{aligned}
\lb{bcfy}
y(0)=f(0),\qqq y'(0)=Q'(0)f(0)+f'(0),\\
y(1)=\r(1)f(1),\qqq y'(1)=Q'(1)\r(1)f(1)+\r(1)f'(1).
\end{aligned}
\]
 The identities \er{bcfy} yield the relations between the boundary conditions for $f$ for $\D_{q}$ and $y$ for $S_p$:
\[
\lb{if2a}
\ca f'(0)-af(0)=0,\\
 f'(1)+b f(1)=0,\ac
 \ \Leftrightarrow \
  \ca y'(0)-(a+Q'(0))y(0)=0,\\
 y'(1)+(b-Q'(1)) y(1)=0,\ac \
 a,b\in \R\cup \{\iy\}.
\]

We consider the eigenvalue problems for $-\Delta_q$ and $S_p$ on
$(0,1)$ subject to  \er{if2a}. Our second main
theorem asserts that the above transformation $- \Delta_q \to S_p$ preserves the
boundary conditions and spectral data.


\begin{theorem}
\lb{TSD} Let $p=G(q),  q\in \mW_1^0$, be defined by \er{dR}. Then
the operators $S_p$ and $-\D_q$,  subject to
the boundary condition \er{if2}, are unitarily equivalent. Moreover, they  have the same
eigenvalues and the norming constants.
\end{theorem}

{\bf Proof.} Let $p=G(q), q\in \mW_1^0$, be defined by \er{dR}. Then
under the transformation $y=\mU f=\r f$ the operators $S_p$ and
$-\D_q$ are unitarily equivalent. Moreover, due to  $y=\r f$ and
\er{if2} the operators $S_p$ and $-\D_q$ have the boundary conditions
given by \er{if2a}.
Using \er{bcfy} we can define the same norming constants. \BBox

\bigskip
Assume that the mapping $p\to $ (eigenvalues + norming constants for
the operator $S_p$) gives the solution of the inverse problem for
the operator $S_p$. Then, since the mapping $p\to q$ is an analytic
isomorphism we obtain  that the solution of the inverse problem for
the mapping  $p\to $ (eigenvalues + norming constants  for the
operator $-\D_q$).

\medskip
Similar arguments work for the operator $-\D_q$ and the associated  inverse problem. We will give a more precise explanation in the proof of Theorems \ref{T1} $\sim$ \ref{T3}.

\medskip
Therefore, the inverse problem for $-\Delta_q$ is solvable if and
only if so is for $S_p$. In this section, we consider the case $E_1=0$, $\nu=1$.


\subsection{Robin boundary condition}
Consider the operator $-\Delta_qf=-{1\/\r^2}(\r^2f')'$ subject to
the boundary condition (\ref{if2}) for the case $a, b\in \R$. We consider
the case $q_0\in \R$ and $E_1=0$. Let $A=(a,b,q_0)\in \R^3$. Let
$\m_n=\m_n(q,A), n\geq 0,$ be the eigenvalues of $\Delta_q$. Then we
have
$$
\m_n(q,A)=\m_n^0+c_0+\wt\m_n(q,A),\quad \mathrm{where} \quad
(\wt\m_n)_{1}^{\iy}\in\ell^2,\qq c_0=\|q\|^2,
$$
and $\m_n^0=(\pi n)^2+2(a+b)$, $n\ge 0$, denote the unperturbed
eigenvalues. We introduce the norming constants
\[
\label{ncab}
\f_n(q,A)=\log\left|\r(1)f_n(1,q,A)\/f_n(0,q,A)\right|,\qquad n\ge
0,
\]
where $f_n$ is the $n$-th eigenfunction. Note that $f_n(1,q,A)\ne 0$
and $f_n(0,q,A)\ne 0$. The inverse problem for $S_p$ with Robin
boundary condition was solved in \cite{KC09}. Therefore, applying
Theorem \ref{T4} and the result of the inverse problem for $S_p$
\cite{KC09}, we have the following theorem.

\begin{theorem}
\label{Tipq3} Let $E_1=0$ for $\n=1$. For each $A=(a,b, q_0)\in\R^3$,
the mapping
$$
\P_{A}:q\mapsto \left((\wt\m_{n}(q,A))_{n=1}^{\iy}\,;
(\f_{n}(q,A))_{n=1}^{\iy}\right)
$$
is a real-analytic isomorphism between $\mW_1^0$ and $\cM_1\ts
\el2_1$, where $\cM_1$ is given by \er{S1Mj} with $\m_n^0=(\pi
n)^2+2(a+b), n\ge 1$.
\end{theorem}
{\bf Proof.} Let $q\in \mW_0^1$ and $A=(a,b,q_0)\in\R^3$. We
consider the Sturm-Liouville problem with the generic boundary
conditions,
$$
-{1\/\r^2}(\r^2f')'=\l f,\qqq f'(0)-af(0)=0,\qqq f'(1)+b f(1)=0.
$$
Let $\m_n=\m_n(q,A), n=0,1,2,...$ be the eigenvalues of the
Sturm-Liouville problem. It is well known that
$$
\m_n(q,A)=\m_n^0+c_0+\wt\m_n(q,A),\quad \mathrm{where}
\quad (\wt\m_n)_{1}^{\iy}\in\ell^2,\qq c_0=\|q\|^2.
$$
Following \cite{KC09}, we introduce the norming constants
\[
\label{ncb}
\f_n(q,A)=\log\left|\r(1)f_n(1,q,A)\/f_n(0,q,A)\right|,\qquad n\ge
0,
\]
where $f_n$ is the $n$-th eigenfunction. Thus for fixed $A\in \R^3$
we have the mapping
$$
\P_{A}:q\mapsto \P_{A}(q)=
\left((\wt\m_n(q,A))_{n=1}^{\iy}\,;(\f_n(q,A))_{n=1}^{\iy}\right)
$$

Let $p=G(q), q\in \mW_0^1$. We use Theorem \ref{TSD}. Consider the
Sturm-Liouville problem
$$
\begin{aligned}
S_p y=-y''+p(x)y,\qquad y'(0)-a_0y(0)=0,\qquad y'(1)+b_0y(1)=0,\\
a,b,q_0\in \R,\qq  a_0=a-q_0,\qq b_0=b+q_0.
\end{aligned}
$$
Denote by $\s_n=\s_n(p), n\ge 0$ the eigenvalues of $S_p$ and let
$\vk_n(p)$ be the corresponding norming constants given by
\[
\label{NuDef}
\vk_n(p)=\log\lt|{y_n(1,p,a_0,b_0)\/y_n'(0,p,a_0,b_0)}\rt|\,,\qquad
n\ge 0.
\]
Recall that  due to \cite{KC09}  (see Prosition 5.4 in \cite{KC09})
for each $a_0,b_0\in\R$  the mapping
\[
\lb{FKC} \F_{a_0,b_0}:p\mapsto \F_{a_0,b_0}(p)=
\left((\wt\s_{n}(p))_{n=1}^{\iy}\,;(\vk_n(p))_{n=1}^{\iy}\right)
\]
is a real-analytic isomorphism between $\mH_0$ and $\cM_1\ts\el2_1$.

Due to Theorem \ref{T4} we obtain the identity
$$
\F_{a_0,b_0}(G(q))=\P_{A}(q),\qq \forall \ q\in \mW_0^1.
$$
The mapping $\P_{A}(\cdot)$ is the composition of two mappings $\F_{a_0,b_0}$ and
$G$, where each of them is the corresponding analytic isomorphism
(see \er{FKC} and Theorem\ref{T4}). Then for each $A\in \R^3$ the
mapping
$$
\P_{A}:q\mapsto
\left((\wt\m_n(q,A))_{n=1}^{\iy}\,;(\f_n(q,A))_{n=1}^{\iy}\right)
$$
is a real-analytic isomorphism between $\mW_0^1$ and
$\cM_1\ts\ell^2_1$. \BBox

\subsection{Dirichlet boundary condition}

On the
interval $[0,1]$ we consider the operator   $-\D_\n=-{1\/\r^2}(\r^2f')'$  with Dirichlet boundary condition. We consider the
case $\n=1$, $q_0\in \R$ and $E_1=0$. Denote by $\m_n=\m_n(q),
n=1,2\cdots$, the eigenvalues of $-\D_1$.  It is well-known that all
$\m_n$ are simple and satisfy
\[
\lb{DBc0} \m_n=\m_n^0+c_0+\wt\m_n,\qq   \m_n^0 = (n\pi)^2, \quad
(\wt\m_n)_{1}^{\iy}\in\ell^2, \qqq c_0=\int_0^1(q_0+q)^2dx,
\]
where $\m_n^0=(\pi n)^2$, $n\ge 1$, are the eigenvalues for the  unperturbed
case $r=1$. We introduce the norming constants
\[
\lb{DBc1} \vk_n(q)=\log\left|\r(1)f_n'(1,q)\/f_n'(0,q)\right|,
\qquad n\ge 1,
\]
where $f_n$ is the $n$-th eigenfunction of $-\Delta_{\nu}$. Note
that $f_n'(0)\ne 0$ and  $f_n'(1)\ne 0$. The inverse problem for
$S_p$ with the Dirichlet  boundary condition was solved in
\cite{PT87}. Therefore, applying Theorem \ref{T4} and the result of
the inverse problem for $S_p$ \cite{PT87}, we have the following
theorem.

\begin{theorem}
\label{Tipq1}
Let $\n=1$ and $E_1=0$. For any $q_0\in\R$ the mapping
$$
\P : q\mapsto \left((\wt\m_{n}(q))_{n=1}^{\iy}\,;
(\vk_{n}(q))_{n=1}^{\iy}\right)
$$
is a real-analytic isomorphism between $\mW_1^0$ and $\cM_1\ts
\el2_1$, where $\cM_1$  is given by \er{S1Mj} with $\m_n^0=(\pi
n)^2, n\ge 1$. In particular, in the symmetric case the spectral
mapping
\[
\wt\m: \mW_0^{1,odd}\to \cM_1,\qqq {\rm given \ by }   \qqq q\to
\wt\m
\]
is a real  real analytic isomorphism between the Hilbert space
$\mW_0^{1,odd}$ and  $\cM_1$.
\end{theorem}

{\bf Proof}. The proof repeats the proof of Theorem \ref {Tipq3},
based on Theorem \ref{TSD} and the well-known results from
\cite{PT87}. \BBox

\subsection{Mixed boundary condition}
We consider the operator  $-\D_\n=-{1\/\rho^2}(\rho^2f')'$ with
mixed boundary condition $f(0)=0,\quad f'(1)+b f(1)=0$, where
$(b,q)\in \R\ts\mW_1^0.$
 We consider the
case $\n=1$, $q_0\in \R$ and $E_1=0$.
 Let $\mu_n=\mu_n(q,b), n=0,1,2,...$ be the associated eigenvalues. They satisfy
\[
\lb{ipC22x}
\begin{split}
 \mu_n(q,b)=\mu_n^0+c_0+\wt\mu_n(q,b), \qq (\wt\mu_n)_{1}^{\iy}\in\ell^2,\qq
c_0=\int_0^1(q_0+q)^2dx.
\end{split}
\]
where $\mu_n^0= \pi^2(n+{1\/2})^2+2b$  are the eigenvalues for for
the unperturbed case $r=1$. We introduce the norming constants
\[
\label{ipC23x}
\c_n(q,b)=\log\left|\r(1)f_n(1,q,b)\/f_n'(0,q,b)\right|,\qquad n\ge
0,
\]
where $f_n$ is the $n$-th  eigenfunction satisfying $f_n'(0,q,b)\ne
0$
 and $f_n(1,q,b)\ne 0$.  When $q =b = 0$, a simple calculation gives
$\c_n^0:=\c_n(0,0)=-\log \pi(n\!+\!{\textstyle{1\/2}}).
$

\begin{theorem}
\lb{Tipq2}
Let $\n=1$ and $E_1=0$ and let $b,q_0\in\R$.
Consider  the inverse problem for \er{ipC21} $\sim$ \er{ipC24} for
any fixed $(b,q_0)\in \R^2$.

(i) The mapping
$$
\P:q\mapsto
\left((\wt\mu_n(q,b))_{n=1}^{\iy}\,;(\c_{n-1}(q,b)-\c_{n-1}^0)_{n=1}^{\iy}\right)
$$
is a real-analytic isomorphism between $\mW_1^0$ and $\mathcal
M_1\ts\ell^2_1$, where $\cM_1$ is given by \er{S1Mj} with
$\m_n^0=\pi^2(n+{1\/2})^2+2b, n\ge 1$.

(ii) For each $(q;b)\in \mW_1^0\ts\R$ the following identity holds true:
\[
\label{IdentityB}
b=\sum_{n=0}^{+\iy} \lt(2-{e^{\c_n(q,b)}\/|{\pa w\/\pa \l}(\mu_n,q,b)|}\rt),
\]
where the function $w(\l,q,b)$ is given by
\[
\label{adamz} w(\l,q,b)=
\cos\sqrt\l\cdot\prod_{n=0}^{+\iy}{\l-\mu_n(q,b)\/\l-\mu_n^0}\,,\qquad \l\in\C.
\]
Here both the product and  the series converge uniformly on bounded
subsets on the complex plane.
\end{theorem}
{\bf Proof}. The proof is based on Theorem \ref{TSD} and the results
from \cite{KC09}. We omit one, since it repeats the proof of Theorem
\ref {Tipq3}.  \BBox


\subsection{Inverse problem for the curvature.}
We define the simple unitary transformation $\mU$ by
 $$
 \mU:
L^2([0,1],rdx)\to L^2([0,1],dx),\qqq y=\mU f= r^{{1\/2}}f, \qq \r=r^{{1\/2}}.
$$

{\bf Proof of Theorem \ref{T5}.} Consider  the inverse problem for
\er{ipGc1}-\er{ipGc3} for fixed $A=(a,b, q_0)\in\R^3$.

i) Let $q_0=0, \n\ge 1$. We have two mappings $\x=G(q)$ and
$$
q\to \P_{A_0}(q)=\rt((\wt\m_{n}(q,A_0))_{n=1}^{\iy}\,;
(\f_{n}(q,A_0))_{n=1}^{\iy}\rt)
$$
and the composition of these mappings
\[
\lb{PG1} \x\to \P_{A_0}(G^{-1}(\x))=\P_{A_0}\circ G^{-1}(\x)
\]
Then due to Theorems \ref{T2} and \ref{T4}, we deduce that the
mapping $\P_{A_0}\circ G^{-1}$ is a real-analytic isomorphism
between $\mH_0$ and $\cM_1\ts \el2_1$, where $\cM_1$ is given by
\er{S1Mj}  with $\m_n^0=(\pi n)^2+2(a+b)$.

ii) Let $q_0\in \R, \n=1,\ E_1=0$ and $(a,b,q)\in \R^2\ts\mW_1^0$.

Consider the Sturm-Liouville operator $-\D_q$ given by
\[
\lb{ipC31aa}
\begin{aligned}
-\D_qf=-{1\/\r^2}(\r^2f')',\qqq f'(0)-af(0)=0,\qquad f'(1)+b f(1)=0.
\end{aligned}
\]
 Let $\m_n=\m_n(q,a,b), n=0,1,2,...$ be the eigenvalues of the
Sturm-Liouville problem \er{ipC31aa}. It is well known that
\[
\lb{ipC32} \m_n=\m_n^0+c_0+\wt\m_n(q,a,b),\quad
\mathrm{where} \quad (\wt\m_n)_{1}^{\iy}\in\ell^2,\qq c_0=\|q\|^2.
\]
Here  $\m_n^0=(\pi n)^2+2(a+b)$, $n\ge 1$ are the unperturbed eigenvalues
for $r=1$. We introduce the norming constants
\[
\label{ipC33}
\f_n(q,a,b)=\log\left|\r(1)f_n(1,q,a,b)\/f_n(0,q,a,b)\right|,\qquad
n\ge 0,
\]
where $f_n$ is the $n$-th eigenfunction. Note that $f_n(1,a,q,b)\ne
0$ and $f_n(0,q,a,b)\ne 0$.

Under the transformation $\mU: L^2([0,1],\r^2dx)\to L^2([0,1],dx)$,
given by $y=\mU f= \r f$, we obtain
$$
\mU (-\D_{\r,u}) \mU^{-1}=S_p+c_0,\qq S_p y=-y''+py,
$$
where due to \er{bcfy} the function $y$ satisfies the following
boundary conditions
\[
\lb{if2x}
\ca f'(0)-af(0)=0,\\
 f'(1)+b f(1)=0,\ac
 \qq \Leftrightarrow \qq
  \ca y'(0)-(a+q_0)y(0)=0,\\
 y'(1)+(b-q_0) y(1)=0,\ac \qq
 a,b\in \R\cup \{\iy\}.
\]

We have two mappings $\x=G(q)$ and
$$
q\to \P_{a,b}(q)=\rt((\wt\m_{n}(q,a,b))_{n=1}^{\iy}\,;
(\f_{n}(q,a,b))_{n=1}^{\iy}\rt)
$$
and the composition of these mappings
\[
\lb{PG1x} \x\to \P_{a,b}(G^{-1}(\x))=\P_{a,b}\circ G^{-1}(\x)
\]
Then due to Theorems \ref{T2} and \ref{T4}, we deduce that the
mapping $\P_{a,b}\circ G^{-1}$ is a real-analytic isomorphism
between $\mH_0$ and $\cM_1\ts \el2_1$, where $\cM_1$ is given by
\er{S1Mj}.
\BBox


\section {Spectral data mapping for the case $q_0=0$}
\setcounter{equation}{0}


\subsection {Non-linear mapping}

In view of (\ref{if1}), we take $\r(x)$ as follows
\[
\lb{dro} \r(x)=e^{Q(x)},\qqq Q=\int_0^xq(t)dt,\qqq q_0=0.
\]
We assume that the potential $u=u(Q)$ is related to $q$ is the
following way.

\medskip
\noindent
\no  {\bf Condition U.} {\it
The function $u(\cdot)$ is real analytic and  satisfies
\[
\lb{con1}
u'(t)\le 0, \qqq \forall \  t\in \R.
\]
\[
\lb{con2}
  \|u'(Q)\|\le F( \|q\|),  \quad  q\in \mW_1^0,
\]
for some increasing function $F: [0, \iy ) \to [0, \iy )$. Here
$\|\cdot\|$ denotes the norm of $L^2(0,1)$.}

\medskip
Since $\r, u$ are related with $q$ by \er{dro} and the Condition U,
we write $\Delta_q$ instead of $\Delta_{\r,u}$. Now, we recall the
theorem from \cite{IK13} about the following  mapping
\[
\lb{Pu}
p=P(q)=q'+q^2+u(Q)-c_0,\qqq  c_0=\int_0^1(q'+q^2+u(Q))dx.
\]

\begin{theorem}
\lb{TPg}
The mapping $P:\mW_1^0\to \mH_0$ given by \er{Pu} is a real analytic
isomorphism between the Hilbert spaces $\mW_1^0$ and $\mH_0$. In particular, the  operator ${\pa P\/\pa q}$ has an inverse for each $q\in
\mW_1^0$. Moreover, it has the following properties.

\noindent
(1) Let $p=P(q), q\in \mW_0^1$. Then  the following  estimates hold true
\[
\begin{aligned}
\lb{eP1}
&\|q'\|^2\le \|p\|^2 \le \|q'\|^2+2\|q^2\|^2+2\|u\|^2-c_0^2,\\
&\|u\|\le \|q\| F(\|q\|).
\end{aligned}
\]
\noindent
(2) The mapping $P(q)-q' : \mW_1^0 \to \mH_0$ is compact.

Furthermore, the mapping $q\to p=P(q), q\in \mW_1^{0,odd}$
given by \er{Pu} is a real analytic
isomorphism between the Hilbert spaces $\mW_1^{0,odd}$ and $\mH_0^{even}$.

\end{theorem}

{\bf Remark.} 1)  The mapping $q\to p=q'+q^2+u-c_0 : \mH_1 \to
\mH_0$ was considered in \cite{K02}. In some cases the mapping
$\mH_0$ into  $\mH_{-1}$ is also useful (see \cite{K03},
\cite{BKK03}).

2) In the case of inverse spectral theory for surfaces of
revolution, we  study the case of the function $u=E\r^{-{4\/d}}$.
 Here $d+1\ge 2$ is the dimension of the surface of revolution and $E\ge 0$
 is a constant.

\

Our second main theorem asserts that the mapping in Theorem
\ref{TPg}
 preserves the boundary conditions and spectral data.

\begin{theorem}
\lb{T2g} Let $p=P(q),  q\in \mW_1^0$, be defined by \er{Pu}. Then
the operators $S_p$ and $\D_q$ are unitarily equivalent. In
particular, they
 have the same boundary conditions, eigenvalues and the norming constants.
\end{theorem}

Therefore, the inverse problem for $\Delta_q$ is solvable if and only
if so is for $S_p$. Let us consider the  following three cases separately.

\subsection{Dirichlet boundary condition : $a=b=\iy$.}
Consider the Sturm-Liouville operator  $\D_{q}$ defined in
  $L^2((0,1);\r^2(x)dx)$, where $\r(x)=e^{Q(x)}$, having the form
$ \D_{q} f=-{1\/ \r^2}(\r^2f')'+u(Q) f$ equipped with the boundary
condition $f(0)=f(1)=0$. Here $Q(x)=\int_0^xq(t)dt,\qqq q_0=0$ and
$u$ satisfies Condition U.

Denote by $\m_n=\m_n(q), n\ge 1$, the eigenvalues of $\D_q$ subject
 to the boundary condition $f(0)=f(1)=0$ for the case $a =b =\infty$.
   It is well-known that all $\m_n$ are simple and satisfy
$$
\m_n=\m_n^0+c_0+\wt\m_n,\quad {\rm where}\quad
(\wt\m_n)_{1}^{+\iy}\in\ell^2, \qq c_0=\int_0^1(q^2+u)dt,
$$
where $\m_n^0=(\pi n)^2$, $n\ge 1$, denote the unperturbed eigenvalues.
The norming constants are defined by
\[
\label{nc00} \vk_n(q)=\log\left|\r(1)f_n'(1,q)\/f_n'(0,q)\right|,\qquad n\ge 1,
\]
where $f_n$ is the $n$-th eigenfunction. Note that $f_n'(0)\ne 0$ and
$f_n'(1)\ne 0$.
We  recall theorem from \cite{IK13}.

\begin{theorem}
\label{Tip1g} Let $a=b=\iy$.
Then  the mapping
$$
\P : q\mapsto \left((\wt\m_{n}(q))_{n=1}^{\iy}\,;
(\vk_{n}(q))_{n=1}^{\iy}\right)
$$
is a real-analytic isomorphism between $\mW_1^0$ and $\cM_1\ts \el2_1$,
where $\mathcal M_1$ is
defined by (\ref{S1Mj}) with $\m_n^0=(\pi n)^2,
n\ge 1$.
In particular, in the symmetric case the spectral mapping
\[
\wt\m: \mW_1^{0,odd}\to \cM_1,\qqq {\rm given \ by }   \qqq p\to
\wt\m
\]
is a real  real analytic isomorphism between the Hilbert space
$\mW_1^{0,odd}$ and  $\cM_1$.

\end{theorem}

\subsection{Mixed boundary condition : $a=\iy, b\in \R$.}
Consider the Sturm-Liouville operator  $\D_{q}$ defined in
  $L^2((0,1);\r^2(x)dx)$, where $\r(x)=e^{Q(x)}>0$, having the form
$ \D_{q} f=-{1\/ \r^2}(\r^2f')'+u(Q) f$ equipped with the mixed
boundary condition $f(0)=0, f'(1)+bf(1)=0$. Here $Q=\int_0^xq(t)dt,\
q_0=0$ and $u$ satisfies Condition U.

Let $\mu_n=\mu_n(q,b), n\geq 0$, be the eigenvalues of $-\D_q$ subject
to the boundary condition $f(0)=0, f'(1)+bf(1)=0$ for the case
$a=\iy, b\in \R$. We then have
$$
\mu_n=\mu_n^0+c_0+\wt\mu_n(q,b),\quad \mathrm{where} \quad
(\wt\mu_n)_{1}^{\iy}\in\ell^2,\qq c_0=\int_0^1(q^2+u)dt,
$$
and $\mu_n^0=\pi^2(n+{1\/2})^2+2b$, $n\ge 0$, denote the unperturbed eigenvalues.
The norming constants are defined by
\[
\label{ncbxx} \c_n(q,b)=\log\left|\r(1)f_n(1,q,b)\/f_n'(0,q,b)\right|,
\qquad n\ge 0,
\]
where $f_n$ is the $n$-th  eigenfunction.
Note that $f_n'(0,q,b)\ne 0$ and $f_n(1,q,b)\ne 0$.
 A simple calculation gives
$$
\c_n^0=\c_n(0,0)=-\log \pi(n\!+\!{\textstyle{1\/2}}),\qquad {\rm where}\qquad
\sqrt{\mu_n^0}=\pi(n\!+\!{\textstyle{1\/2}}).
$$
We  recall theorem from \cite{IK13}.

\begin{theorem}
\lb{Tip2g}  i) For each fixed
$b\in \R$ the mapping
$$
\P:q\mapsto
\left((\wt\mu_n(q,b))_{n=1}^{\iy}\,;(\c_{n-1}(q,b)-\c_{n-1}^0)_{n=1}^{\iy}\right)
$$
is a real-analytic isomorphism between $\mW_1^0$ and $\mathcal M_1\ts\ell^2_1$,
where $\mathcal M_1$ is
defined by (\ref{S1Mj}) with $\m_n^0=\pi^2(n+{1\/2})^2+2b,
n\ge 1$.

ii) For each $(q,b)\in \mW_0^1\ts\R$ the following identity holds true:
\[
\label{IdentityBa}
b=\sum_{n=0}^{\iy} \lt(2-{e^{\c_n(q,b)}\/|{\pa w\/\pa \l}(\mu_n,q,b)|}\rt),
\]
where
\[
\label{adamy} w(\l,q,b)=
\cos\sqrt\l\cdot\prod_{n=0}^{+\iy}{\l-\mu_n(q,b)\/\l-\mu_n^0}\,,\qquad \l\in\C.
\]
Here both the product and  the series converge uniformly on bounded subsets
on the complex plane.

\end{theorem}

\subsection{Robin boundary condition : $a,b\in \R$.}
Consider the Sturm-Liouville operator  $\D_{q}$ defined in
  $L^2((0,1);\r^2(x)dx)$, where $\r(x)=e^{Q(x)}>0$, having the form
$ \D_{q} f=-{1\/ \r^2}(\r^2f')'+u(Q) f$ equipped with the generic
boundary condition $f'(0)-af(0)=0, f'(1)+bf(1)=0$. Here
$Q=\int_0^xq(t)dt,\ q_0=0$ and $u$ satisfies Condition U.

Let $\m_n=\m_n(q,a,b), n\geq 0,$ be the eigenvalues of $\Delta_q$
subject to the boundary condition $f'(0)-af(0)=0, f'(1)+bf(1)=0$ for
the case $a, b\in \R$. Then we have
$$
\m_n=\m_n^0+c_0+\wt\m_n(q,a,b),\quad \mathrm{where} \quad
(\wt\m_n)_{1}^{\iy}\in\ell^2,\qq c_0=\int_0^1(q^2+u)dt,
$$
and $\m_n^0=(\pi n)^2+2(a+b)$ denote the unperturbed eigenvalues.
We introduce the norming constants
\[
\label{ncab4}
\f_n(q,a,b)=\log\left|\r(1)f_n(1,q,a,b)\/f_n(0,q,a,b)\right|, \qquad
n\ge 1,
\]
where $f_n$ is the $n$-th eigenfunction. Note that $f_n(1,a,q,b)\ne
0$ and $f_n(0,q,a,b)\ne 0$. We recall the results from \cite{IK13}.

\begin{theorem}
\label{Tip3g} For any $a,b\in\R$, the mapping
$$
\P_{a,b}:q\mapsto \left((\wt\m_{n}(q,a,b))_{n=1}^{\iy}\,;
(\f_{n}(q,a,b))_{n=1}^{\iy}\right)
$$
is a real-analytic isomorphism between $\mW_1^0$ and $\cM_1\ts \el2_1$, where
$\cM_1$ is given by
 (\ref{S1Mj}) with $\m_n^0=(\pi  n)^2+2(a+b)$.
\end{theorem}


\subsection{Proof of Theorems \ref{T1} $\sim$\ref{T3}.}
Recall that due to \er{3}  we obtain that the Laplacian on $(M, g)$
is unitarily equivalent to a direct sum of one-dimensional
Schr\"odinger operators, namely, $ \lb{} -\D_{(M,g)}\backsimeq
\os_{\n\ge 1} \D_\n, $ where the direct sum acts in $\os_{\n\ge 1}
L^2([0,1],dx)$. We consider the inverse problem for the operator
$\D_\n$ for fixed $\n\ge 1$ and $q_0=0$.

 {\bf Proof of Theorem \ref{T1}.}  We consider
 the inverse problem
for the operator $\D_q$ given by
\[
\lb{snH}
\begin{aligned}
&\D_\n =-{1\/\r^2} \pa_x \r^2 \pa_x +{E_\n\/r^2} ,\\
&\r=r^{m\/2}=\r_0e^{Q}, \qq Q(x)=\int_0^x(q_0+q)dt,\qq q\in \mW_1^0,
\end{aligned}
\]
under the Dirichlet boundary conditions $f(0)=f(1)=0$ and for each
$\n\ge 1$.

Consider the case $q_0=0$. We apply Theorem \ref{Tip1g} to our
operator $\D_\n$, since the function $u={E_\n\/r^2}=E_\n
e^{-{4\/m}Q}$ satisfies Condition U. Then Theorem \ref{Tip1g} gives
that the mapping $ \P : q\mapsto \left((\wt\m_{n}(q))_{1}^{\iy}\,;
(\vk_{n}(q))_{1}^{\iy}\right) $ is a real-analytic isomorphism
between $\mW_1^0$ and $\cM_1\ts \el2_1$, where $\cM_1$ is given by
\er{S1Mj} with $\m_n^0=(\pi n)^2$. In particular, in the symmetric
case the spectral mapping $\wt\m: \mW_0^{1,odd}\to \cM_1$ given by $
q\to \wt\m $ is a real real analytic isomorphism between the Hilbert
space $\mW_0^{1,odd}$ and  $\cM_1$.

The case $\n=1$ and $E_1=0$ has been considered in Theorem \ref{Tipq1}.
\BBox

\bigskip

{\bf Proof of Theorem \ref{T2}.}
 We consider the inverse problem for the operator $-\D_\n$  givn by
 \er{snH},
under the mixed boundary conditions $f(0)=0, f'(1)+bf(1)=0$
for any fixed $(b,\n)\in \R\ts\N$.

Consider the case $q_0=0$. We apply Theorem \ref{Tip2g} to our
operator $-\D_\n$, since the function $u={E_\n\/r^2}=E_\n
e^{-{4\/m}Q}$ satisfies Condition U. Then Theorem \ref{Tip2g} gives
that the mapping
$$
\P:q\mapsto
\left((\wt\mu_n(q,b))_{n=1}^{\iy}\,;(\c_{n-1}(q,b)-\c_{n-1}^0)_{n=1}^{\iy}\right)
$$
is a real-analytic isomorphism between $\mW_1^0$ and $\mathcal M_1\ts\ell^2_1$,
 where
$\mathcal M_1$ is given by \er{S1Mj} with $\m_n^0=(\pi
n+{1\/2})^2+2b$. Moreover, for each $(q;b)\in \mW_1^0\ts\R$ the
following identity holds true:
\[
\label{IdentityBb}
b=\sum_{n=0}^{+\iy} \lt(2-{e^{\c_n(q,b)}\/|{\pa w\/\pa \l}(\mu_n,q,b)|}\rt),
\]
where the function $w(\l,q,b)$ is given by
\[
\label{adamx} w(\l,q,b)=
\cos\sqrt\l\cdot\prod_{n=0}^{+\iy}{\l-\mu_n(q,b)\/\l-\mu_n^0}\,,\qquad \l\in\C.
\]
where both the product and  the series converge uniformly on bounded subsets
on the complex plane.

The case $\n=1$ and $E_1=0$ has been considered in Theorem \ref{Tipq2}.
\BBox

{\bf Proof of Theorem \ref{T3}.} We consider the inverse problem for
the operator $-\D_\n$   given by\er{snH}, under the generic boundary
conditions $f'(0)-af(0)=0, f'(1)+bf(1)=0$ for any fixed $(a,b,\n)\in
\R^2\ts\N$.

Consider the case $q_0=0$. We apply Theorem \ref{Tip3g} to our
operator $-\D_\n$, since the function $u={E_\n\/r^2}=E_\n
e^{-{4\/m}Q}$ satisfies Condition U. Then Theorem \ref{Tip3g} gives
that the mapping
$$
\P_{a,b}:q\mapsto \left((\wt\m_{n}(q,a,b))_{n=1}^{\iy}\,;
(\f_{n}(q,a,b))_{n=1}^{\iy}\right)
$$
is a real-analytic isomorphism between $\mW_1^0$ and $\cM_1\ts
\el2_1$, where $\cM_1$ is given by \er{S1Mj} with $\m_n^0=(\pi
n)^2+2(a+b)$.

The case $\n=1$ and $E_1=0$ has been considered in Theorem \ref{Tipq3}.
\BBox

\setlength{\itemsep}{-\parskip} {\footnotesize \no  {\bf
Acknowledgments.} Various parts of this paper were written during
Evgeny Korotyaev's stay in the Mathematical Institute of University
of Tsukuba, Japan and Mittag-Leffler Institute, Sweden. He is
grateful to the institutes for the hospitality. His study was
supported by the RSF grant  No. 15-11-30007. }

\end{document}